\documentclass[%
 reprint,
 amsmath,amssymb,
 aps,
]{revtex4-1}

\usepackage{graphicx}
\usepackage{dcolumn}
\usepackage{bm}
\usepackage{color}

\newcommand{\avg}[1]{\langle#1\rangle}

\begin{document}


\title{Epidemic Spreading on Activity-Driven Networks with Attractiveness}%

\author{Iacopo Pozzana}
 \affiliation{Birkbeck Institute for Data Analytics - Birkbeck, University of London - London, UK}
 \author{Kaiyuan Sun}
 \affiliation{Laboratory for the Modeling of Biological and Socio-technical Systems, Northeastern University, Boston, USA}
\author{Nicola Perra}%
 \email{n.perra@greenwich.ac.uk}
\affiliation{Centre for Business Network Analysis, Greenwich University, London, UK
}%

\date{\today}

\begin{abstract}
We study SIS epidemic spreading processes unfolding on a recent generalisation of the activity-driven modelling framework. In this model of time-varying networks each node is described by two variables: activity and attractiveness. The first, describes the propensity to form connections. The second, defines the propensity to attract them. We derive analytically the epidemic threshold considering the timescale driving the evolution of contacts and the contagion as comparable. The solutions are general and hold for any joint distribution of activity and attractiveness. The theoretical picture is confirmed via large-scale numerical simulations performed considering heterogeneous distributions and different correlations between the two variables. We find that heterogeneous distributions of attractiveness alter the contagion process. In particular, in case of uncorrelated and positive correlations between the two variables, heterogeneous attractiveness facilitates the spreading. On the contrary, negative correlations between activity and attractiveness hamper the spreading. The results presented contribute to the understanding of the dynamical properties of time-varying networks and their effects on contagion phenomena unfolding on their fabric.

\begin{description}
\item[PACS numbers]
89.75.-k, 64.60.aq, 87.23.Ge
\end{description}
\end{abstract}
\maketitle


Many social, natural and technological systems can be modelled as networks. The structure of such systems is often not fixed and exhibits complex temporal dynamics~\cite{holme2015modern, holme12,masuda2016guide,porter2016dynamical}.
However, the large majority of studies revolve around representations that neglect the role of time~\cite{barrat08, boccaletti06,newman2010networks}. In particular, connections are typically approximated as either static or annealed~\cite{vespignani2012modelling,boguna2013nature}. 
Since networks are often used as an environment for the study of dynamical processes, the choice concerning which approximation to adopt is a matter of time scales: when the process is faster than the network evolution, the network structure can be assumed static; in the opposite conditions, networks can be effectively described by annealed representations.
When, however, the time scale of the process studied is comparable
to the one characterising the network evolution, static or annealed approximations
are not viable and can lead to
incorrect conclusions  such as misrepresenting i) the spreading potential of a disease~\cite{morris93-1,morris07-1,valdano15,prakash2010virus,alex12-1,Rocha:2010,Bajardi:2011,vanhems2013estimating,Stehle:2011nx,perra12a,takaguchi2013bursty,holme2014birth,Karsai:2011,toro2007,liu13-1,holme2015basic,wang2016statistical,sun15,han2015epidemic,rizzo2016network}, ii) the exploring capabilities of random walkers~\cite{starnini12,starnini_rw_temp_nets,ribeiro12-2,perra12-2,lambiotte12-1,masuda2016random}, iii) the features of social interactions~\cite{karsai13-1,clauset07,Isella:2011,pfitzner13-1,starnini13-2,takaguchi12-1,gonccalves2015social,fournet2014contact,barrat2015face,sekara2016fundamental,holme03,marton12,ubaldi16,ubaldi2017burstiness}, or the processes of iv) information spreading~\cite{Miritello:2011,dynnetkaski2011,panisson11-1,weng2013role,gleeson2016effects}, v) synchronisation~\cite{albert2011sync}, vi) percolation~\cite{Parshani:2010}, vii) consensus~\cite{consensus_temporal_nrets_2012}, viii) competition~\cite{artime2017dynamics}, ix) social contagion~\cite{liu2017social}, and x) innovation~\cite{rizzo2016innovation}.

Thanks to the unprecedented availability of large and longitudinal datasets,
in recent years a great effort has been put into the development of temporal network representations and models. See References~\cite{holme2015modern, holme12,masuda2016guide} for detailed reviews on the subject. 

One proposal for an analytical model of temporal network comes from the \emph{activity-driven model}~\cite{perra12a}, which relates the temporal structure of the connections to one fundamental
quantity, the \emph{activity}. This feature represents the propensity of a node to establish connections per unit time. In the model, each node $i$ is equipped with an activity $a_i$ extracted from a distribution
$F(a)$. At any time step $t$, nodes are active and thus willing to establish connections with probability proportional to their activity. 
One praise of this simple mechanism is that it relates the contact dynamics to the structure of the time-integrated network:
the resulting degree distribution $P(k)$ depends on the form of $F(a)$,
and in particular a power-law distributed activity produces a power-law degree distribution \cite{perra12a}.
This fact is particularly significant in relation with social networks, which are known to
exhibit distributions of this kind both for the degree \cite{barabasi2016network,barrat08} and for the activity \cite{perra12a,ribeiro13,ubaldi16,karsai14,tomasello2014role}.

In its original form, the activity-driven
model is extremely simple, thus relatively lightweight for performing calculations. Nonetheless it gives rise to a non-trivial temporal structure having an impact
on the unfolding of dynamical processes \cite{perra12a,perra12b,ribeiro13,liu14}.
Precisely because of its simplicity, and in particular of its reliance on only one
node property (the activity), the original activity-driven is not able to reproduce other widespread properties
of social networks, namely finite clustering, assortative mixing, a bursty contact
sequence and memory effects~\cite{granovetter73, newman03, barabasi05}.
 For this reasons, and also thanks to its flexibility,
modifications to the original model have been introduced and investigated
\cite{karsai14, laurent15, sun15, moinet15, ubaldi16,ubaldi2017burstiness}. 

Here, we consider a recent extension of the model~\cite{alessandretti17-1} in which beyond the activity distribution,
the network is characterised by an \emph{attractiveness} distribution~\cite{alessandretti17-1}.This feature accounts for the fact that
some nodes may be a preferential target of interactions (i.e.\ be more popular), in the same way activity
accounts for the fact that some nodes are more inclined to be their initiators.
The attractiveness $b_i$ of a node $i$ is a measure of its propensity to attract contacts. Therefore it is to some extent the reciprocal of the activity, and a natural complement
to it within the model. Heterogeneous attractiveness distributions have been observed in different networks such as online communities \cite{ghoshal06, valverde07, alessandretti17-1}, face-to-face interactions~\cite{starnini13a}, and animal hierarchies \cite{sapolsky05}.

In Ref.~\cite{alessandretti17-1}, besides the introduction of the model, the authors studied the effects of the attractiveness
and of its interplay with the activity on the fundamental dynamical processes: the random walk. Here, we keep investigating how attractiveness and activity affect spreading processes on temporal networks focusing on contagion phenomena. In particular, we consider SIS epidemic models in the case of a generic joint activity-attractiveness distribution deriving an analytic expression for the epidemic threshold. We give a detailed treatment of three scenarios.
First, we examine the case of uncorrelated activity and attractiveness. Second, inspired by observations on real data~\cite{alessandretti17-1}, we study the case of positive correlation between the two variables. Finally, we complete our analysis considering the case of negative correlation. In all cases we use numerical simulations to validate our results.

The shape of the activity distribution, and its second moment in particular,
has been shown to influence the unfolding of different kinds of processes unfolding on activity-driven networks \cite{perra12a, perra12b, starnini14, liu14,sun15,starnini13b}. In case of spreading phenomena, the more heterogeneous the activity distribution
(i.e.\ the larger its variance),
the easier it is for a disease to reach a finite portion of the network \cite{perra12a, starnini14, liu14, sun15}.
Here, we found how the presence of a heterogeneous attractiveness has an
analogous impact. The presence of positive correlations between activity and attractiveness further facilitates the contagion,
while the presence of negative correlations, conversely, hinders it.

The paper is structured as follows:
in Section \ref{model} we present the model and we discuss the attractiveness and its correlation with the activity; in Section \ref{epi} we study the epidemic threshold for an SIS process in the case of absence of activity-attractiveness correlation (\ref{unc}), and in the case of deterministic correlation (\ref{det}), treating all cases both analytically and with simulations; Section \ref{conc} contains the conclusions and an address of possible future works.

\section{THE MODEL} \label{model}

In the original activity-driven model (AD), a network of $N$ nodes is characterised
by an activity distribution $F(a)$ from which the activity, $a_i$, of each node $i$ is extracted.
The model uses a discrete time framework, with time steps of duration
$\Delta t$. At the beginning of each time step, a node $i$ may activate; the
activation happens with probability $a_i \Delta t$; if a node activates,
it will form a fixed number $m$ of connections towards randomly selected nodes
(multiple connections, as well as self-connections, are forbidden, and in general $m \ll N$); the connections
remain active for the duration of the time step, at the end of which they are all
reset, and the process starts again.

The above depicted is the original version of the model, as proposed
in \cite{perra12a}. There, when a node activates, it will choose
the targets of its connections among all other members of the network with
equal probability. In the version of the model we consider here, which we will call
\emph{activity-driven with attractiveness} (ADA), the network is
also characterised by an attractiveness distribution~\cite{alessandretti17-1}. In general,
the two values of activity $a_i$ and attractiveness $b_i$ for the same
node $i$ are not necessarily uncorrelated, and are sampled from a joint  probability distribution
$H(a,b)$. Interestingly, recent observations on online social networks have shown both variables to behave according to a power-law with similar exponents and an approximately linear correlation \cite{alessandretti17-1}.  

The ADA works like the AD, except that when a node $i$ activates it will choose another
node $j$ as a target of one of its $m$ connections with probability proportional to the
second node's attractiveness, $b_j$. As the probability of choosing any node
must be equal to one, the correct normalisation for the probability
is $b_j/\avg{b} N$, where $\avg{b}$ is the
mean value of the attractiveness:
\begin{equation}
\avg{b} = \int da\, db\, H(a,b)\, b.
\end{equation}
The model thus behaves similarly to a \emph{linear preferential attachment},
as the overall number of contacts received by a node during any time-window is linearly proportional to its attractiveness; the total number of contacts (received and initiated), on the other hand, depends on both activity and attractiveness.

By time-integrating the connections obtained at different time steps we can study the emergent topological properties of the network. In particular, the time-integrated network over $T$ time steps is defined as the union of the instantaneous networks obtained at $T$ different time steps,
i.e.\ two nodes figure as connected in the integrated network if they are connected in at least one of the $T$ instantaneous networks.
A weight, equal to the number of instantaneous networks in which the edge appears, can also be associated to each edge.
For the AD, the degree distribution of the integrated network is connected to the activity distribution through the relation $P(k) \sim F(k/Tm)$, as long as it holds $k \ll T \ll N$ \cite{perra12a}. The study of the time-integrated properties of ADA networks will be instead matter of future work.

\begin{figure}
 \centering
  {\includegraphics[width=0.99\columnwidth]{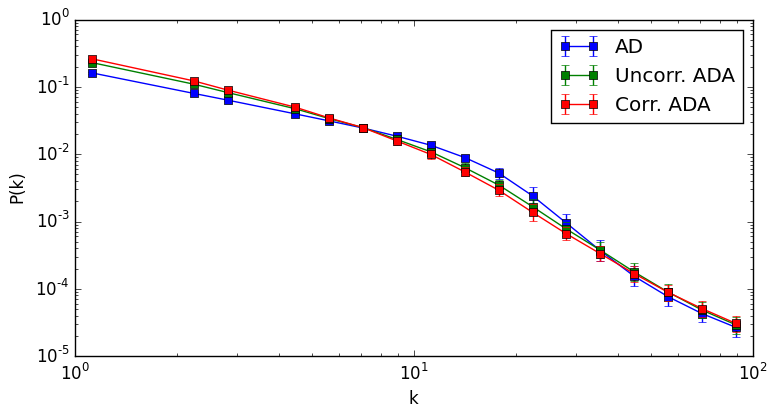}}
  {\includegraphics[width=0.99\columnwidth]{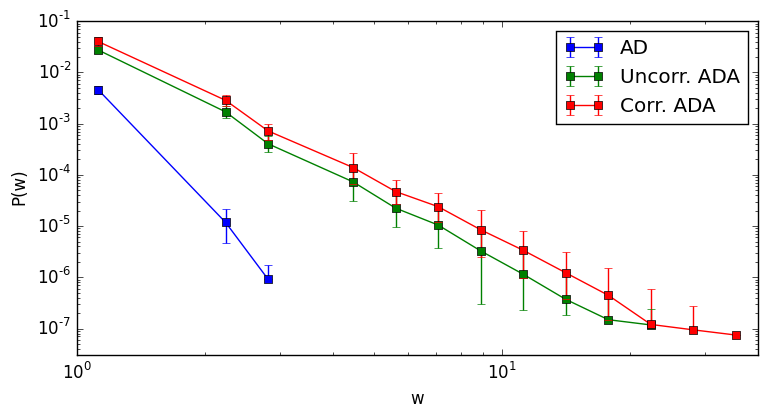}}
  \caption[degree distributions]
  {
  Degree and edge weight distribution for time-integrated ADA and AD. Both distributions show long tails. We used an activity distribution $F(a) \propto a^{-2.4}$ for all the three networks.
  For the uncorrelated ADA, the attractiveness distribution is also $G(b) \propto b^{-2.4}$.
  For the correlated case, we set $b_i \propto a_i$ for each node $i$.
  We run our simulations on networks of $10^5$ nodes with $m = 5$, integrated over $10^3$ time steps and averaged over $100$ runs.
  }
  \label{deg}
\end{figure}

FIG. \ref{deg} illustrates the degree distribution $P(k)$ and the edge weight distribution $P(w)$ obtained
for two ADA and an AD networks of size $N= 10^5$ after a time-integration of $T=10^3$ time steps.
We used an activity distribution $F(a) \propto a^{-2.4}$.
In the AD model all nodes have equal attractiveness; in the uncorrelated ADA we used an attractiveness distribution
independent on, but identical to, the activity distribution: $G(b) \propto a^{-2.4}$; 
for the correlated ADA we set the attractiveness of every node to be proportional to its activity:
$b_i \propto a_i,\, i = 1, \dots, N$.
The exponents are chosen to be representative of typical values encountered in social networks \cite{newman03, boccaletti06}.
The plot of the degree distribution shows that the presence of heterogeneous attractiveness in ADA networks does not induce dramatic changes.
However, the inspection of the weight distribution highlights the difference between the two models.
The presence of heterogeneous attractiveness induces heterogeneity in the partner selection
that reverberate in the weight distribution. As we will see later, such heterogeneity favours the contagion process.

The ADA model differs from recent generalisations of AD networks introduced in \cite{karsai14} and further expanded in \cite{laurent15, ubaldi16,ubaldi2017burstiness}. In fact, in these extensions local reinforcement mechanisms have been used as a way to model the emergence and evolution of strong/weak ties~\cite{granovetter73}. However, local mechanisms alone cannot explain the dynamics of ties especially in the current social media landscape where people can easily be in contact with celebrities or access information provided by popular accounts. The attractiveness describes scenarios of \emph{global popularity}, as opposed to cases of \emph{local} reinforcement where
the perceived attractiveness of a node may change between its peers, so that the contact probability is encoded in pairs rather than in the single nodes;
also, we model the attractiveness as constant in time, not being strengthened nor weakened by the occurrence of contacts - or its lack.

\section{EPIDEMIC THRESHOLD} \label{epi}

As we discussed above, the presence of a heterogenous attractiveness affects the temporal structure of contacts.
We want to quantify this phenomenon by studying its impact
on a dynamical process; namely, we choose to evaluate the epidemic threshold
for an SIS process. The fact that the analytical value of such threshold has
already been calculated and tested for the original activity-driven in \cite{perra12a}
allows us to straightforwardly draw a comparison between the AD and the ADA.

The SIS is an example of a compartmental epidemic model \cite{kermac27-1,keeling08-1,barrat08};
in this framework, every node belongs to a
certain class with respect to the disease status: susceptible (S), or infected (I). When a susceptible node contacts (or is contacted by)
an infected one, it may become itself infected, with probability $\lambda$.
Meanwhile, infected nodes can undergo spontaneous recovery with rate $\mu$ and become susceptible again.

In general contagion processes are characterised by a threshold which determines whether the disease is able to spread in the system affecting a macroscopic fraction of nodes~\cite{kermac27-1,keeling08-1,barrat08,pastor2015epidemic, wang2016statistical}. 
In the limit of static networks the epidemic threshold of a SIS processes is determined by the spectral properties of the adjacency matrix~\cite{wang03}. In the limit of annealed networks and uncorrelated topologies the threshold is defined by the moments of the degree distribution~\cite{pastor01a}. Interestingly, a closed expression for the threshold of a SIS process unfolding on a general time-varying networks has been obtained~\cite{prakash2010virus, valdano15}. This can be expressed in terms of the spectral properties of the \emph{system-matrix} which is defined as $\mathbf{S}=\prod_t [(1-\mu)\mathbf{A_t}+\lambda]$ (where $A_t$ is the adjacency matrix at time $t$). Despite its generality, this expression hides the physics of the process behind the computation of eigenvalues which is typically done numerically.

The AD framework allows an explicit mathematical derivation~\cite{perra12a}.  In particular, the  threshold, for SIS models, depends on the moments of
the activity distribution:
\begin{equation} \label{ad_thresh}
 \frac{\lambda}{\mu} \avg{k} > \frac{2\avg{a}}{\avg{a} + \sqrt{\avg{a^2}}}
 \equiv T_{AD},
\end{equation}
at the first order in $N^{-1}$ and activity~\cite{perra12a}.
We have introduced $T_{AD}$,
that denotes the value of the epidemic threshold for the activity-driven model and depends on the properties the network, which in turn are determined by the moments of the activity distribution.

It is important to stress how in the expression of the threshold the time integrated properties of the network (as the degree distribution) do not appear. The dynamical properties are defined only by the activity distribution.  

Here, we extend the literature providing an explicit analytical expression for the epidemic threshold for an SIS process in the ADA model for any form of the probability distribution $H(a,b)$.
To do so, we assume nodes to be
characterised by their activity and attractiveness values alone,
and accordingly group them in classes; nodes within each class are considered
statistically equivalent (\emph{mean-field assumption}). We also assume that the two variables are discretely distributed,
but the derivation would apply as well to the case of continuous variables.
We denote with $N_{a,b}$ the number of nodes
of activity $a$ and attractiveness $b$, with the condition
$\sum_{a,b} N_{a,b} = N$. The number of susceptible and infected nodes
of activity $a$ and attractiveness $b$ at time $t$ is indicated as $S_{a,b} (t)$
and $I_{a,b}$ respectively.
A master equation for the temporal evolution of the number of infected nodes in each class
can be written, again, in the limit of large size $N \gg 1$, where the probability
of having repeated contacts between the same two nodes can be neglected.
Without lack of generality in the following we will set $\Delta t = 1$. The master equation reads:
\begin{equation}
    \begin{split}
        &I_{a,b}(t+1) = (1 - \mu) I_{a,b}(t) + \\
        &\frac{\lambda m}{N\avg{b}} S_{a,b}(t) \Bigl[ a \sum_{a',b'} b'I_{a',b'}(t) +
        b \sum_{a',b'} a'I_{a',b'}(t) \Bigr].
    \end{split}
\end{equation}
The first term on the right side
accounts for the infected inherited from the previous time step, minus
the cases of spontaneous recovery. The two terms in bracket represent, the first,
the probability for a susceptible node in the class $(a,b)$ to activate and contact an
infected node in any other class, and the second term represents the probability for an infected node
in any other class to activate and contact a susceptible node in the class $(a,b)$; the
difference with the AD model is that now the probability for a node in the class
$(a,b)$ to be contacted depends on $b$, and the probability for it to contact a node in
the class $(a',b')$ depends on $b'$.
We can define two auxiliary functions to simplify what follows:
\begin{gather}
    \theta(t) \equiv \sum_{a,b} a I_{a,b}(t),\\
    \phi(t) \equiv \sum_{a,b} b I_{a,b}(t).
\end{gather}

In considering the initial phase of the spreading, when $I_{a,b}(t) \ll N_{a,b}$,
we can take $S_{a,b}(t) \simeq N_{a,b}$;
the master equation becomes:
\begin{equation}
 I_{a,b}(t+1) \simeq (1 - \mu) I_{a,b}(t) + \frac{\lambda m}{N\avg{b}} N_{a,b} [ a \phi(t) + b \theta(t) ].
\end{equation}
From the last one, we can obtain three more equations: one by summing aver all classes,and two more by first multiplying by $a$ and $b$ respectively, and then summing.

Switching to a continuous time regime, we obtain a system of three linear differential equations:
\begin{gather}
 \frac{\partial I(t)}{\partial t} \simeq -\mu I(t) + \frac{\lambda m}{\avg{b}} [\avg{a} \phi(t) + \avg{b} \theta(t)], \\
 \frac{\partial \theta(t)}{\partial t} \simeq -\mu \theta(t) + \frac{\lambda m}{\avg{b}} [\avg{a^2} \phi(t) + \avg{ab} \theta(t)], \\
 \frac{\partial \phi(t)}{\partial t} \simeq -\mu \phi(t) + \frac{\lambda m}{\avg{b}} [\avg{ab} \phi(t) + \avg{b^2} \theta(t)],
\end{gather}
of eigenvalues:
\begin{equation}
 \kappa_0 = -\mu, \quad
 \kappa_\pm = \frac{\lambda m}{\avg{b}} \left(\avg{ab} \pm \sqrt{\avg{a^2}\avg{b^2}} \right) - \mu.
\end{equation}

The outbreak prevails when at least one eigenvalue is positive.
The $\kappa_{\pm}$ recover the eigenvalues of the AD if we use a constant attractiveness;
$\kappa_0$ is not a candidate for being the largest eigenvalue, unless $\kappa_0 = \kappa_+$;
so the epidemic threshold is determined by the positivity of $\kappa_+$, leading to:
\begin{equation} \label{ada_thresh}
 \frac{\beta}{\mu} >
 \frac{2\avg{a}\avg{b}}{\avg{ab} + \sqrt{\avg{a^2}\avg{b^2}}} \equiv T_{ADA},
\end{equation}
where we have used $\avg{k} = 2m\avg{a}$~\cite{perra12a}, and introduced $\beta \equiv \lambda \avg{k}$ as the per capita spreading rate.
As for the AD (above), we use $T_{ADA}$ to denote the value of the epidemic threshold encoded in the features of nodes activity and attractiveness.

Eq.~\ref{ada_thresh} is valid for any form of $H(a,b)$;
in particular, the expression recovers the value from Eq.~\ref{ad_thresh} if a constant attractiveness is used,
so that $\avg{b^2} = \avg{b}^2$ and $\avg{ab} = \avg{a}\avg{b}$.
In the reminder of this section we focus
on two scenarios: first, when activity and attractiveness
are  uncorrelated; second, when they are instead deterministically
correlated - either positively on negatively.

To provide a precise estimation of the threshold value when simulating an SIS process,
we use the lifetime-based method introduced in \cite{boguna2013nature} and also used in \cite{sun15} for the same purpose.
The lifetime $L$ is defined as the amount of time elapsed before the disease either dies out or spreads to a finite fraction $C$ of the network.
The rationale behind this definition is the following:
when the system is below threshold, the disease is not able to spread and dies out in a short time;
above threshold, the lifetime is also short, as the disease can quickly reach a finite fraction of nodes.
Only for values of $\beta/\mu$ close to the threshold we expect to observe a longer lifetime,
as the contrasting effects of contagion and spontaneous recovery
are almost equally strong and they struggle to prevail on each other;
we can thus take the value of $\beta/\mu$ corresponding to the maximum in $L$ as an estimation of $T_{ADA}$. Indeed, the life time, obtained by averaging
over many realisations, is equivalent to the susceptibility in standard percolation theory~\cite{boguna2013nature}, thus it provides a precise method to detect the threshold numerically.

The peak in the lifetime is more pronounced for larger values of $N$
as, for power-law activity and attractiveness distributions,
the heterogeneity effects lowering the epidemic threshold are
constrained by the finite size of the system \cite{satorras02finite}.
For such reason, we run our simulations on networks of increasing size and expect to observe the peak increase and occur
at lower values of $\beta/\mu$, converging
to the theoretical value as $N \to \infty$.
In particular, here we chose to use $N$ equal to $10^4$,
$10^5$ and $10^6$; we also set $C = 0.25$ as the target
value for the network coverage (fraction of infected nodes).

\subsection{Uncorrelated distributions}
\label{unc}

In the absence of correlations, $H(a,b)$ can be written as a product
$H(a,b) = F(a) G(b)$, where $F(a)$ is the activity
distribution and $G(b)$ the attractiveness distribution.
In Eq.~\ref{ada_thresh}, the mean value of the product can be factorised to obtain:
\begin{equation}
\label{threshold11}
T_{ADA} = \frac{2}{1 + \sqrt{\frac{\avg{a^2}\avg{b^2}}{\avg{a}^2\avg{b}^2}}}.
\end{equation}

We can see that, once fixed the average values, the threshold can be made arbitrarily small by increasing
either or both the second moments, i.e.\ introducing heterogeneity.
As the threshold depends on the moments of the two distributions in the same way,
the case with constant attractiveness and generic $F(a)$ (the AD)
can be mapped to the one with constant activity and attractiveness distribution
$F(b)$.

As $\avg{b^2} \geq \avg{b}^2$ always holds, the threshold can only be lower than or equal to the one found in the AD;
this means that the introduction of any amount of heterogeneity in the attractiveness helps the epidemic spreading.

As an example of uncorrelated distributions, let us consider two power-laws: $F(a) = Ca^{-\gamma_a}$ and $G(b) = Db^{-\gamma_b}$; in both cases $a$ and $b$ are bounded inside the interval $[\epsilon,1]$, to avoid divergences.

\begin{figure}
 \centering
  {\includegraphics[width=0.95\columnwidth]{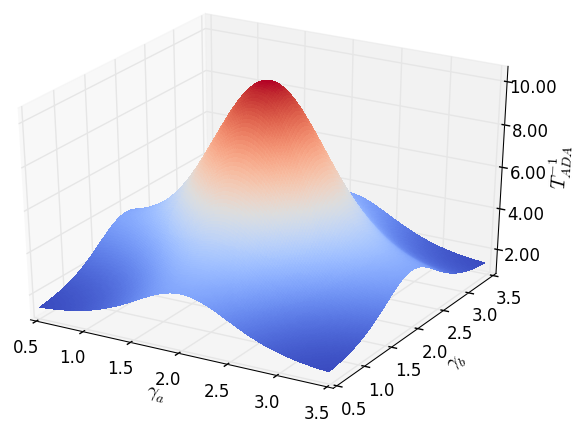}}
  \caption[Epidemic threshold as a function of the exponents]
  {Value of $1/T_{ADA}$ for the case of uncorrelated distributions: $H(a,b) = F(a) G(b)$,
  where both variables are distributed according to a power-law with values in the range $[10^{-3},1]$: $F(a) \propto a^{-\gamma_a}$ and $G(b) \propto b^{-\gamma_b}$.
  The threshold depends on the exponents $\gamma_a$ and $\gamma_b$ via the moments of the two distributions.
  We plot the reciprocal of the epidemic threshold, so that the larger is the value plotted the easier is for the disease to spread.
  $T_{ADA}$ has the same dependence on the two exponents and it is symmetric around its maximum in $\gamma_a = \gamma_b = 2$.}
  \label{3d}
\end{figure}

FIG.~\ref{3d} illustrates the behaviour of the epidemic threshold obtained from Eq.~\ref{threshold11}; we report the values of $T_{ADA}^{-1}$ -
so that the plot shows a maximum when the spreading potential is maximum (the threshold shows a minimum) - as a function of the two exponents $\gamma_a$ and $\gamma_b$.
The threshold exhibits the same dependence on each of the two exponents, as the analytic expression also shows, with local maxima for $\gamma_a = 2$ (and $\gamma_b = 2$) and a global maximum in $\gamma_a = \gamma_b = 2$. The function is symmetric around such value.

\begin{figure}
 \centering
  {\includegraphics[width=0.95\columnwidth]{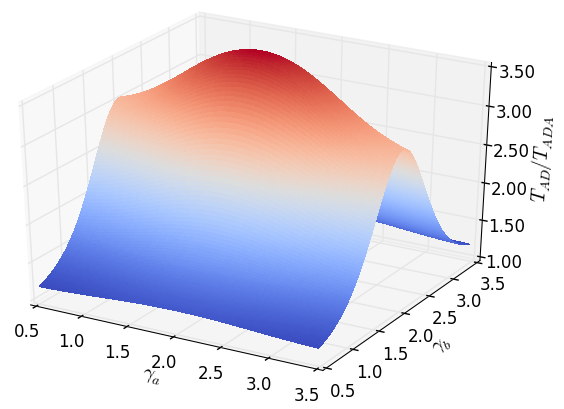}}
  \caption[Comparison between ADA and AD]
  {Plot of $T_{AD}/T_{ADA}$, the ratio of the epidemic threshold in the original model
  and the epidemic threshold with attractiveness,
  for the case of uncorrelated distributions:
  $H(a,b) \propto a^{-\gamma_a} b^{-\gamma_b}$ on the ADA.
  The activity on the AD in distributed according to the same law as on the ADA:
  $F(a) \propto a^{-\gamma_a}$.
  The ratio is a function of the two exponents ($T_{AD}$ depends on $\gamma_a$ only).
  When $\gamma_b$ diverges or tends to zero, the attractiveness distribution loses heterogeneity and the ADA converges to the AD. The difference is maximal for $\gamma_b = 2$.}
 \label{comparison}
\end{figure}

In FIG.~\ref{comparison} we show an analytical comparison between the ADA and the AD, by plotting $T_{AD}/T_{ADA}$ -
the ratio between the epidemic threshold in the original model
and the epidemic threshold with attractiveness computed following the analytical solution -
as a function of the two power-law exponents.
As expected from the equations we find that, for values of $\gamma_b$ either very large or close to zero, the ratio tends to one, as $\avg{b^2} \rightarrow \avg{b}^2$ and consequently $T_{ADA} \rightarrow T_{AD}$.
Otherwise, the attractiveness always lowers the threshold thus facilitating the spreading of the epidemic phenomenon.

\begin{figure}
 \centering
   {\includegraphics[width=0.99\columnwidth]{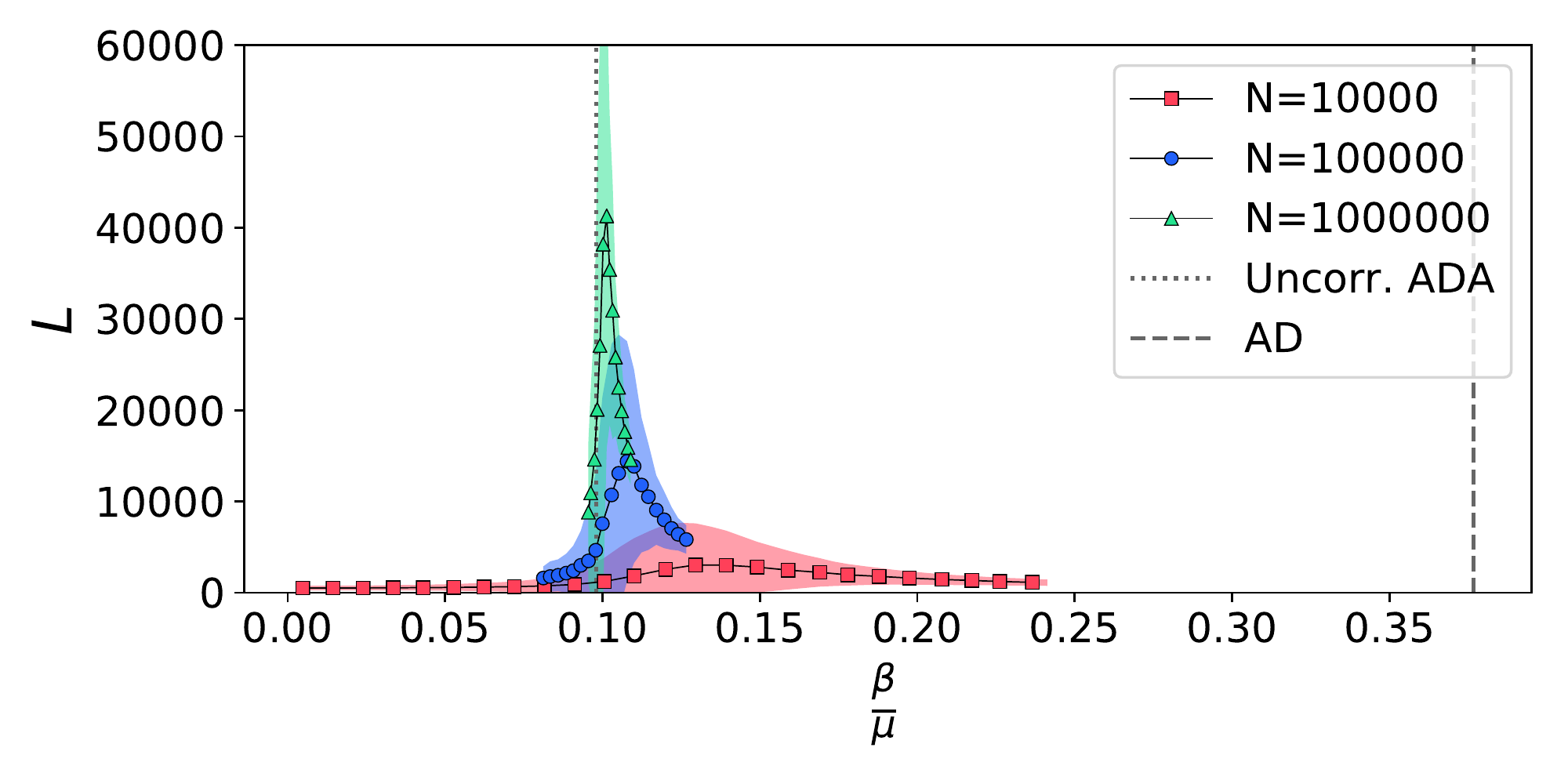}}
  {\includegraphics[width=0.99\columnwidth]{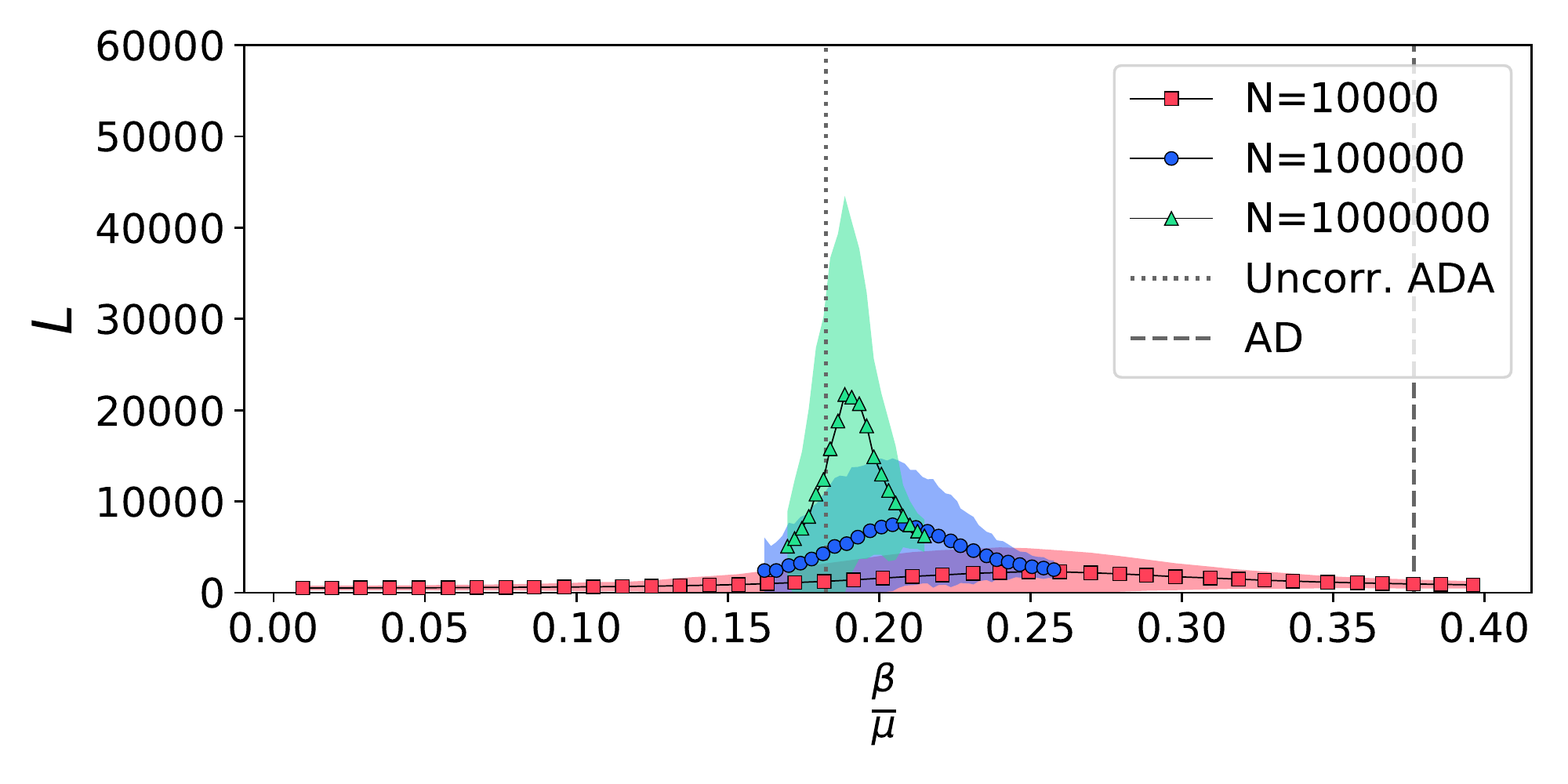}}
  \caption[Power-law activity and attractiveness distributions]
  {Lifetime of SIS processes on an ADA uncorrelated networks with power-law distributed activity ($\gamma_a = 1.8$)
  and attractiveness (top: $\gamma_b = 2.1$, bottom: $\gamma_b = 2.8$), plotted for different values of $\beta/\mu$. We let $\lambda$ vary while we keep fixed $\mu = 0.01$ and $\avg{k}$ is determined by the relation $\avg{k} = 2m \avg{a}$.
  The theoretical epidemic
  threshold (dotted line) is appreciably lower than the threshold for the AD (dashed line), and is in
  accordance with the simulations. We used $m = 2$, $\epsilon =10^{-3}$. In the upper plot, we used:
  for $N=10^4$, $3000$ simulations (red);
  for $N=10^5$, $1000$ simulations (blue);
  for $N=10^6$, $500$ simulations (green).
  In the lower plot, we used:
  for $N=10^4$, $5000$ simulations (red);
  for $N=10^5$, $500$ simulations (blue);
  for $N=10^6$, $500$ simulations (green).
  Solid lines with markers and shaded areas represent mean and $95\%$ confidence interval separately.}
 \label{powlaw}
\end{figure}

We tested the validity of our findings by simulating an SIS process with two
different choices of the exponents: one with $\gamma_a = 1.8$ and $\gamma_b = 2.1$,
the other with same $\gamma_a$ and $\gamma_b = 2.8$;
in both cases we set $m=2$
and took the median value over a number realisations of the process varying between $500$ and $5000$ (depending on the size).
The results are shown in FIG.~\ref{powlaw}, where we plotted the lifetime of the process for different values
of $\beta/\mu$. In particular, we let $\lambda$ vary while we keep fixed $\mu = 0.01$ and $\avg{k}$ also does not change, being determined by the relation $\avg{k} = 2m \avg{a}$.
The lifetime exhibits a peak converging towards the theoretical prediction (dotted line) for increasing values of $N$.
Also, the comparison with the AD (dashed line) shows that the epidemic threshold is appreciably lower in the ADA setting, as the heterogeneity due to the attractiveness distribution facilitates the spreading.

\subsection{Deterministic correlation}
\label{det}

As a second example, we study the case of a deterministic activity-attractiveness correlation, where the value of one variable uniquely
determines the value of the other one for any given node. The joint distribution can be expressed in the form:
\begin{equation}
 H(a,b) = F(a) \delta(b - q(a)),
\end{equation}
where $\delta(x)$ is the Dirac delta and $q(a)$ is the function that determines the attractiveness of a node
given its activity: $b_i = q(a_i),\, \forall i$.
Using the relation $G(b) = F(a) |da/db|$ we can obtain an expression
for $G(b)$ (provided $q(a)$ has an inverse):
\begin{equation}
G(b) = F(q^{-1}(b)) \left| \frac{dq^{-1}(b)}{db} \right|.
\end{equation}

A case of interest is $q(a) \propto a^{\gamma_c},\, \gamma_c > 0$, so that if one the variables is power-law distributed,
the other is too, with different exponent: if for example $F(a) \propto
a^{-\gamma_a}$, then the attractiveness will be distributed as
$G(b) \propto b^{-1+\frac{1-\gamma_a}{\gamma_c}}$. This also includes the case of identical correlation, for $\gamma_c = 1$.
A generic moment of the joint distribution can be expressed as:
\begin{equation}
 \avg{a^n b^m} = \avg{a^{n+\gamma_c m}},
\end{equation}
and Eq.~\ref{ada_thresh} becomes:
\begin{equation}
 T_{ADA} = \frac{2\avg{a}\avg{a^{\gamma_c}}}{\avg{a^{1+\gamma_c}} + \sqrt{\avg{a^2}\avg{a^{2\gamma_c}}}}.
\end{equation}

\begin{figure}
 \centering
  {\includegraphics[width=0.95\columnwidth]{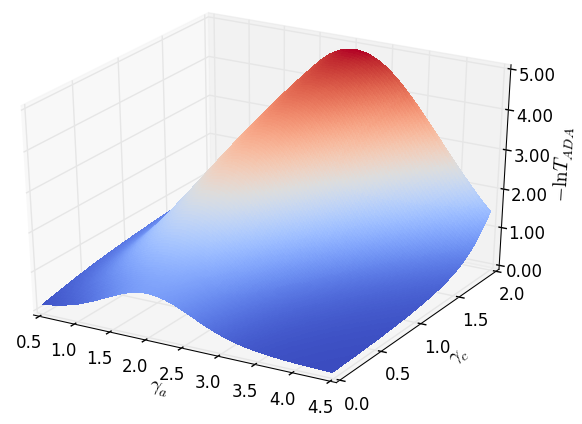}}
  \caption[Deterministic activity-attractiveness correlation]
  {For the case of deterministic activity-attractiveness correlation of the form $b \propto a^{\gamma_c}$,
  with activity distribution $F(a) \propto a^{-\gamma_a}$,
  the plot shows the logarithm of $T_{ADA}^{-1}$ as a function of the two exponents $\gamma_a$ and $\gamma_c$.
  Higher values in the plot correspond to a lower epidemic threshold,
  thus to an easier spreading.
  In particular, when $\gamma_c = 0$ -
  which corresponds to the AD case as the attractiveness is constant,
  the spreading is maximal for $\gamma_a = 2$ as expected. When $\gamma_c$ increases,
  the threshold value decreases very rapidly,
  as the most active nodes also become the most popular ones.}
 \label{3ddet}
\end{figure}

FIG.~\ref{3ddet} shows the behaviour of the threshold as a function
of the exponents $\gamma_a$, governing the activity distribution,
and $\gamma_c$, which determines the activity-attractiveness relation
as depicted above. We report the values of the logarithm of $T_{ADA}^{-1}$:
as in previous plots,
we choose to show the reciprocal of the epidemic threshold. In this case we also choose to take the logarithm,
as the value of $T^{-1}_{ADA}$ changes considerably in the range studied.
For $\gamma_c = 0$, which is equivalent to the AD as $G(b)$ is constant, the threshold shows a maximum for $\gamma_a = 2$ as expected; as $\gamma_c$ increases,
the maximum increases very quickly as the most active nodes become more and more popular,
greatly facilitating the spreading of the disease.

\begin{figure}
 \centering
  {\includegraphics[width=0.99\columnwidth]{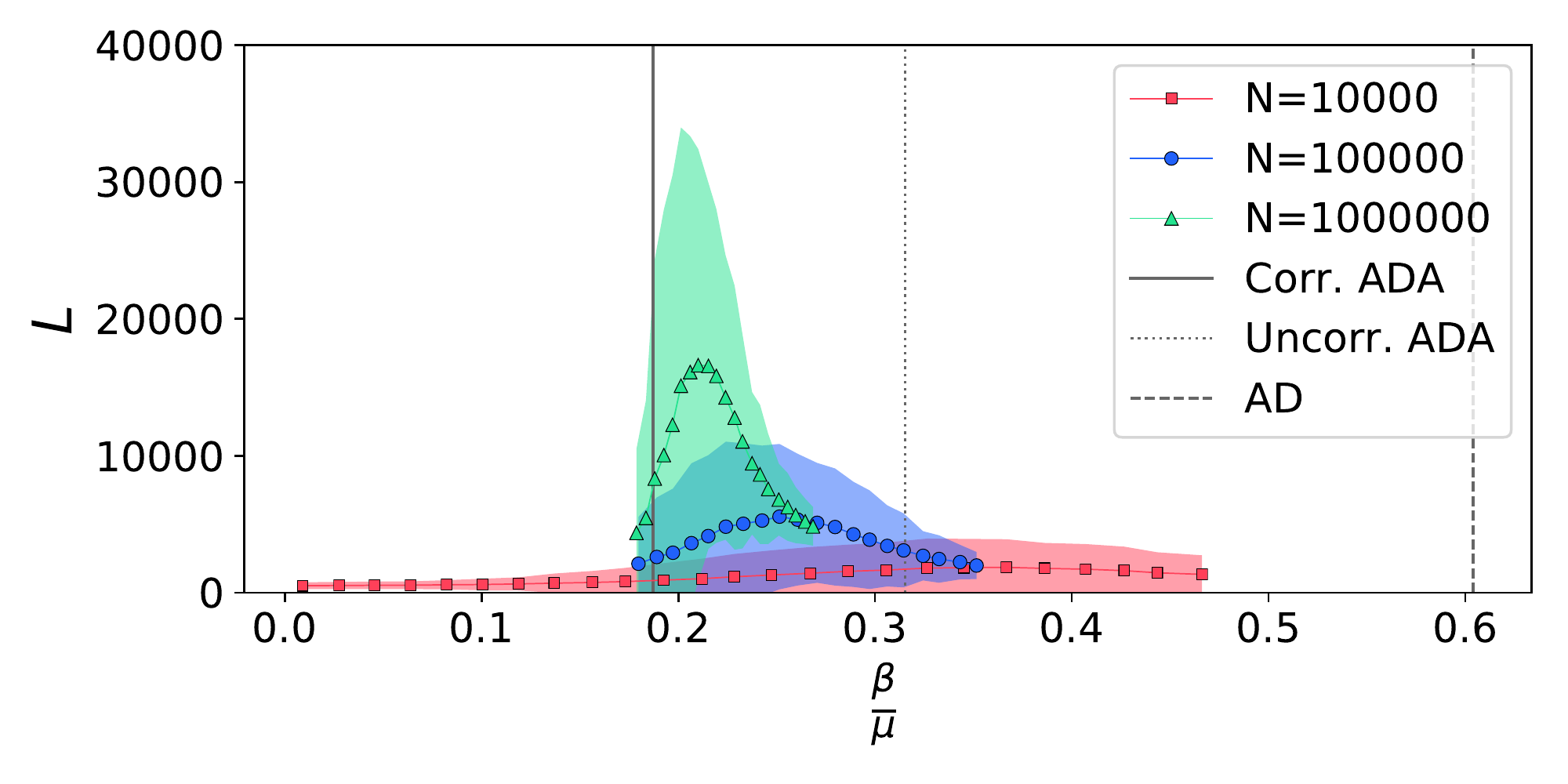}}
  \caption[Identical activity-attractiveness correlation]
  {Lifetime of an SIS process on an ADA network
  with identical activity-attractiveness correlation ($b \propto a$),
  plotted for different values of $\beta/\mu$ and different network sizes;
  we used a power-law distributed activity with $\gamma_a = 2.8$ and let $\lambda$ vary while keeping fixed $\mu = 0.01$ and $\avg{k} = 2m\avg{a}$.
  The theoretical epidemic threshold (solid line) is lower than the threshold for the AD (dashed line). The case of identical but uncorrelated distributions ($\gamma_a = \gamma_b = 2.8$, dotted line)
  is also significantly different from the correlated case.
  We used $m = 2$, $\epsilon =10^{-3}$, and run:
  for $N=10^4$, $3000$ simulations (red);
  for $N=10^5$, $800$ simulations (blue);
  for $N=10^6$, $500$ simulations (green).
  Solid lines with markers and shaded areas represent mean and $95\%$ confidence interval separately.}
 \label{identical}
\end{figure}

We validated the case of identical correlation with computer simulations, by studying an SIS process.
In FIG.~\ref{identical} we plotted the lifetime for different values of $\beta / \mu$.
For increasing values of $N$, the lifetime exhibits a peak that converges towards
the predicted threshold (solid line) -
which is significantly lower than the one obtained in the AD (dashed line), and
also lower than the threshold for uncorrelated distributions (dotted line) - thus confirming our analytical
predictions. We let $\lambda$ vary while keeping fixed $\mu = 0.01$, $\gamma_a = 2.8$, $m = 2$, taking the median over
a number of realisations ranging between $500$ and $3000$, depending on the size of the network.

As a second instance of the deterministically correlated case, we consider $q(a) \propto a^{-1}$.
This form of the function accounts for a case of negative activity-attractiveness correlation,
where the most active nodes are the least attractive and vice versa.
The same formulae hold as for the case of positive $\gamma_c$ above,
leading to the expressions:
\begin{equation}
 \avg{a^n b^m} = \avg{a^{n - m}},
\end{equation}
for a generic moment, and:
\begin{equation}
T_{ADA} = \frac{2}{1 + \sqrt{\frac{\avg{a^2}\avg{a^{-2}}}{\avg{a}^2\avg{a^{-1}}^2}}},
\end{equation}
for the threshold.
In this scenario, opposite to the case of positive correlation considered above,
we expect the correlations to work against the epidemic, as the most active potential spreaders are
also the least attractive and hence the least likely to be infected in the first place.

\begin{figure}
 \centering
  {\includegraphics[width=0.99\columnwidth]{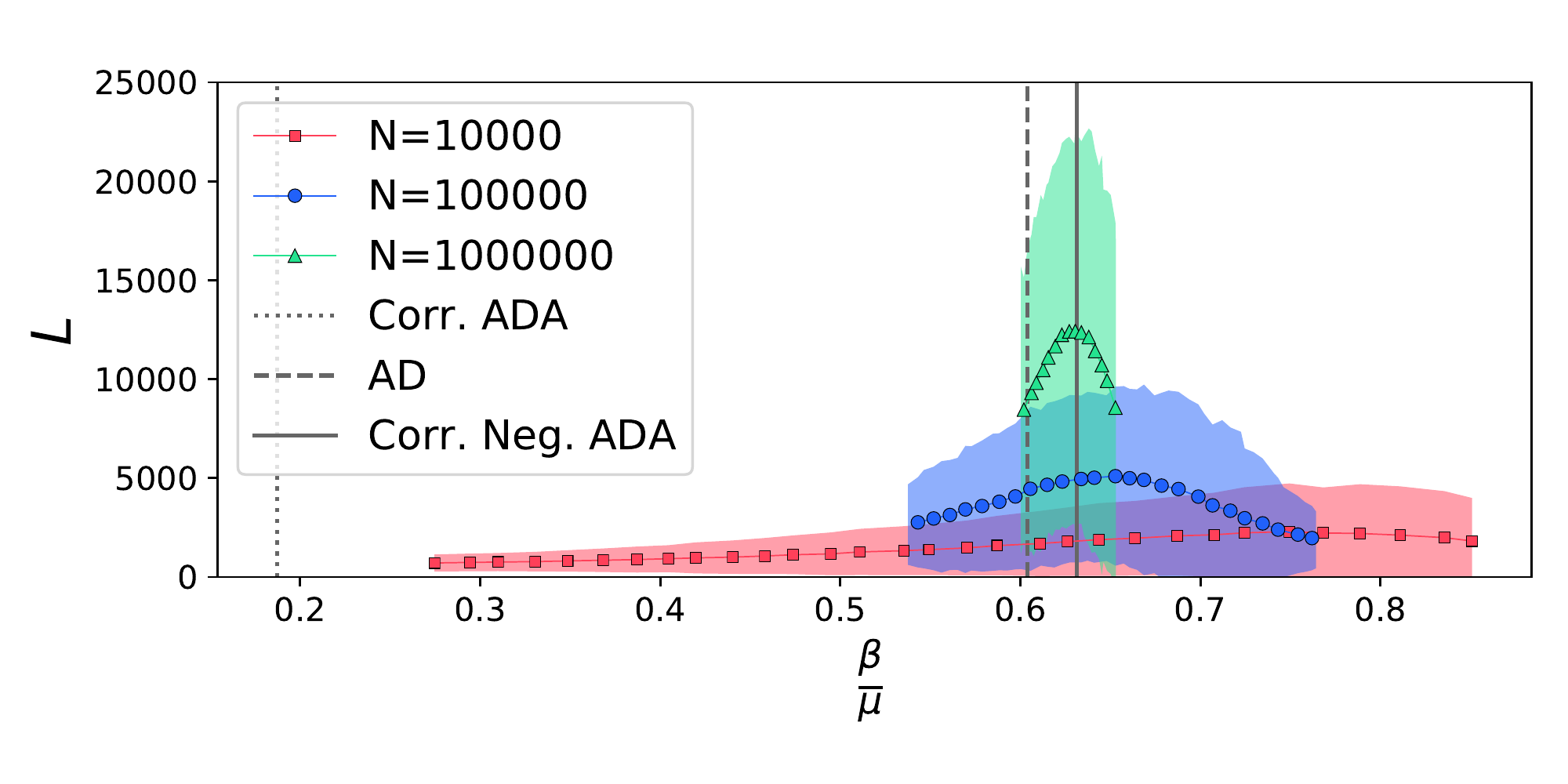}}
  \caption[Negative activity-attractiveness correlation]
  {Lifetime of an SIS process on an ADA network
  with negative deterministic activity-attractiveness correlation of the type $b \propto a^{-1}$,
  plotted for different values of $\beta/\mu$ and different network sizes;
  we used a power-law distributed activity with exponent $\gamma_a = 2.8$ and let $\lambda$ vary while keeping fixed $\mu = 0.01$ and $\avg{k} = 2m\avg{a}$.
  We set $m = 2$, $\epsilon =10^{-3}$, and run:
  for $N=10^4$, $3000$ simulations (red);
  for $N=10^5$, $2000$ simulations (blue);
  for $N=10^6$, $2500$ simulations (green).
  Solid lines with markers and shaded areas represent mean and $95\%$ confidence interval separately.
  The results match the theoretical prediction for the epidemic
  threshold (solid line). The comparison with the case
  of identical correlation shows how the two scenarios
  produce contrasting effects, with a negative correlation hindering the spreading phenomenon.}
 \label{negative}
\end{figure}

Our expectation is corroborated by the experiments described in FIG.\ \ref{negative},
where we study the lifetime of a SIS process on an ADA network characterised by 
an activity-attractiveness correlation of the form $b \propto a^{-1}$ for all nodes,
the activity being distributed as a power-law with exponent $\gamma_a = 2.8$.
We used $m = 2$, $\epsilon = 10^{-3}$ and a
number of realisations ranging from $2000$ to $3000$ for different sizes.
The outcome of the simulations matches well the theoretical threshold (solid line),
and the comparison with the case of identical correlation (i.e.\ same activity
distribution with $\gamma_a = 2.8$, and attractiveness $b \propto a$; dotted line) shows
a stark difference between the two cases, highlighting the contrasting effects of the two
phenomena. The threshold for an analogous AD network (with activity distribution
of power $\gamma_a = 2.8$ and constant attractiveness; dashed line) is close to the negatively correlated case.

\section{CONCLUSIONS}
\label{conc}

We studied a recent generalisation (labelled ADA for simplicity) of the activity-driven model where a second
node's property, called attractiveness, is added. This variable accounts for the fact that
not all nodes are as likely to be the target of interactions initiated by others.
The original activity-driven model (labelled AD for simplicity) would only account for heterogeneity in nodes' behaviour
by distinguishing between more and less active ones, while implicitly assuming constant attractiveness. Observations in different types of real networks show this is not always
the case.

We studied the unfolding of epidemic processes on ADA networks. In particular, we derived analytically an expression for the epidemic threshold of SIS models. The analytical and numerical comparison between spreading dynamics unfolding on ADA and AD networks shows how the introduction of
a new grade of heterogeneity due to a non-constant attractiveness can significantly
alter the spreading of a disease. To precisely quantify the interplay between the activity and attractiveness
we considered three cases.
In the first case we used two power-law uncorrelated distributions.
The results in
this setting show that the introduction of heterogeneity in the attractiveness of nodes facilitates the spreading.
In the second instead, we considered a scenario capturing observations in real networks~\cite{alessandretti17-1}
in which the two quantities follow heterogeneous and correlated distributions.
In this case we found that correlations between the two variables facilitate the spreading process even further.
Finally, we completed our analysis by considering a case of negative correlations;
opposite to the previous case, we found that this type of correlations hinders the spreading phenomenon.

Many of the limits of the AD model are still present in the ADA, i.e.\ absence of high-order correlations, or the absence of burstiness. These properties will be the subject of future extensions of the model. An investigation of the topology of the time-integrated network could also provide some interesting insight, particularly by determining whether the introduction of the attractiveness, and of an appropriate activity-attractiveness correlation, can lead to the emergence of the desired properties that characterise most real social systems (assortativity and clustering). 

Overall, our results contribute to the recent discussion around the effects of temporal connectivity patterns on dynamical processes unfolding on their fabrics. 
 
\bibliography{ada_biblio}

\end{document}